\documentclass[10pt,superscriptaddress,tightenlines,twocolumn,amsmath,amssymb,aps, pra]{revtex4-2}
\UseRawInputEncoding
\usepackage{graphicx}
\usepackage{dcolumn}
\usepackage{bm}
\usepackage{color}
\usepackage{amsmath}
\usepackage{amssymb}
\usepackage{subfigure}
\usepackage{braket}
\usepackage{indentfirst}
\usepackage{mathrsfs}

\usepackage{amssymb}
\usepackage{tcolorbox}
\usepackage{amsmath}
\usepackage{amsfonts}
\usepackage{graphicx}
\usepackage{makecell}
\usepackage{hyperref}
\usepackage{makecell}
\usepackage{color}%
\usepackage{lmodern}
\usepackage{lineno}
\usepackage{relsize}

\usepackage{booktabs}
\usepackage{array}
\usepackage{multirow}
\usepackage[section]{placeins}

\providecommand{\U}[1]{\protect\rule{.1in}{.1in}}

\hypersetup{colorlinks=true, linkcolor=red, citecolor=blue, urlcolor=black}

\begin{document}

\title{Experimental coherent-state quantum secret sharing with finite pulses}
\author{Yuan-Zhuo Wang}\thanks{These authors contributed equally.}
\author{Xiao-Ran Sun}\thanks{These authors contributed equally.}
\author{Xiao-Yu Cao}\thanks{These authors contributed equally.}
\affiliation{National Laboratory of Solid State Microstructures and School of Physics, Collaborative Innovation Center of Advanced Microstrucstures, Nanjing University, Nanjing 210093, China}
\affiliation{Department of Physics and Beijing Key Laboratory of Opto-electronic Functional Materials and Micro-nano Devices, Key Laboratory of Quantum State Construction and Manipulation (Ministry of Education), Renmin University of China, Beijing 100872, China}
\author{Hua-Lei Yin}\email{hlyin@ruc.edu.cn}
\affiliation{Department of Physics and Beijing Key Laboratory of Opto-electronic Functional Materials and Micro-nano Devices, Key Laboratory of Quantum State Construction and Manipulation (Ministry of Education), Renmin University of China, Beijing 100872, China}
\affiliation{National Laboratory of Solid State Microstructures and School of Physics, Collaborative Innovation Center of Advanced Microstrucstures, Nanjing University, Nanjing 210093, China}
\affiliation{Beijing Academy of Quantum Information Sciences, Beijing 100193, China}
\author{Zeng-Bing Chen}\email{zbchen@nju.edu.cn}
\affiliation{National Laboratory of Solid State Microstructures and School of Physics, Collaborative Innovation Center of Advanced Microstrucstures, Nanjing University, Nanjing 210093, China}

\date{\today}

\begin{abstract}
Quantum secret sharing (QSS) plays a significant role in multiparty quantum communication and is a crucial component of future quantum multiparty computing networks. Therefore, it is highly valuable to develop a QSS protocol that offers both information-theoretic security and validation in real optical systems under a finite-key regime. In this work, we propose a three-user QSS protocol based on phase-encoding technology. By adopting symmetric procedures for the two players, our protocol resolves the security loopholes introduced by asymmetric basis choice without prior knowledge of the identity of the malicious player. Kato's concentration inequality is exploited to provide security against coherent attacks with the finite-key effect. Moreover, the practicality of our protocol has been validated under a 30-dB channel loss with a transmission distance of 5-km fiber. Our protocol achieves secure key rates ranging from 432 to 192 bps by choosing different pulse intensities and basis selection probabilities. Offering enhanced security and practicality, our protocol stands as an essential element for the realization of quantum multiparty computing networks.
\end{abstract}

\maketitle

\section{\label{introduction}INTRODUCTION}
Multiparty quantum communication, a key element in realizing quantum multiparty computing networks, offers information-theoretic security that surpasses the limitations of classical cryptography.~\cite{kimbleQuantumInternet2008,gisinQuantumCommunication2007,wehnerQuantumInternetVision2018,bennett2014quantum,ekertQuantumCryptographyBased1991,sasakiPracticalQuantumKey2014,weng2023beating,lucamariniOvercomingRateDistance2018a,yinExperimentalQuantumSecure2023,chen2021integrated,zhou2023experimental,wang2022twin,gu2022experimental,liAllphotonicQuantumRepeater2023,cao2024experimental,azumaAllphotonicQuantumRepeaters2015,frohlichQuantumAccessNetwork2013,jing2024experimental,schiansky2023demonstration}. As a cryptographic component of quantum communication technologies, quantum secret sharing (QSS) has garnered significant interest for its potential in multiparty quantum communication networks~\cite{liBreakingRatedistanceLimitation2023,deoliveiraExperimentalHighdimensionalQuantum2020,williamsQuantumSecretSharing2019,cleveHowShareQuantum1999,hilleryQuantumSecretSharing1999a,fu2015long,zhouQuantumSecretSharing2018,luSecretSharingQuantum2016,bellExperimentalDemonstrationGraphstate2014,weiExperimentalCircularQuantum2013}. In the scenario of secret sharing, each party possesses a portion of the information, with the complete key reconstructed only through collaboration among authorized parties. This is the fundamental concept of the secret sharing scheme, initially introduced independently by Shamir~\cite{shamirHowShareSecret1979} and Blakley~\cite{blakleySafeguardingCryptographicKeys1979}. The security of the classical secret sharing scheme relies on computational complexity, which has been proven to be vulnerable under the rapid development of quantum computing technology~\cite{groverFastQuantumMechanical1996,shorPolynomialTimeAlgorithmsPrime1999}. 

QSS was proposed to address this issue. In 1999, Hillery, Buzek, and Berthiaume utilized the entanglement properties of the Greenberger-Horne-Zeilinger (GHZ) state to construct a QSS protocol~\cite{hilleryQuantumSecretSharing1999a}. This protocol divides a message from a dealer into two parts and distributes them to two players, in such a way that both parts of the split message are necessary to reconstruct the original message. Numerous QSS protocols were proposed in succession~\cite{fu2015long,guoQuantumSecretSharing2003,gaoDeterministicMeasurementdeviceindependentQuantum2020,zhouQuantumSecretSharing2018,dengCircularQuantumSecret2006,markhamGraphStatesQuantum2008,qinCryptanalysisHilleryBuIfmmode2007,gaoDeterministicMeasurementdeviceindependentQuantum2020,yanQuantumSecretSharing2005,dengImprovingSecurityMultiparty2005,xiaoEfficientMultipartyQuantumsecretsharing2004,lanceTripartiteQuantumState2004} while the feasibility of QSS protocols was verified through experiments~\cite{luSecretSharingQuantum2016,bellExperimentalDemonstrationGraphstate2014,weiExperimentalCircularQuantum2013,bogdanskiExperimentalQuantumSecret2008,gaertnerExperimentalDemonstrationFourParty2007,schmidExperimentalSingleQubit2005,chenExperimentalQuantumSecret2005}. Despite the advancements, existing QSS protocols still face challenges in terms of security, efficiency, and practicality. For instance, the original three-user GHZ-based QSS protocol is vulnerable if one player is dishonest~\cite{qinCryptanalysisHilleryBuIfmmode2007}. Over the past few decades, protocols employing sequential single-qubit transmission have been proposed to circumvent the need for GHZ states~\cite{tavakoliSecretSharingSingle2015,hai-qiangExperimentalSingleQubit2013,schmidExperimentalSingleQubit2005}. However, these single-qubit protocols have drawbacks in terms of security, failing to resist Trojan horse attacks~\cite{heCommentExperimentalSingle2007,schmidSchmidReply2007,xuSecureQuantumKey2020a}. In recent years, differential phase-shift QSS schemes have been developed employing weak coherent states to simplify the experimental setup~\cite{guDifferentialPhaseShift2021,jiaDifferentialPhaseShift2021,inoueDifferentialphaseshiftQuantumSecret2008}. However, this kind of QSS protocol exhibits security limitations, offering resistance only against individual attacks. The round-robin QSS schemes have gained popularity in recent years due to their inherent advantages, including high noise tolerance and resistance against signal disturbances~\cite{guSecureQuantumSecret2021,weiQuantumSecretSharing2018}. However, the practical implementation of this scheme is constrained by the requirement of a variable-delay Mach-Zehnder interferometer at the dealer's end. Recently, a QSS protocol encoding logic bits in weak coherent states, which does not require intensity modulation and phase randomization, has been proposed~\cite{shenExperimentalQuantumSecret2023}. However, the protocol exhibits asymmetry in certain steps for different participants, necessitating prior knowledge of the eavesdropper's identity and potentially introducing security loopholes.
\par In this paper, we propose a tripartite QSS protocol. Our protocol utilizes phase encoding with weak coherent states, a method validated in numerous phase-encoded quantum key distribution (QKD) protocols~\cite{loSecurityQuantumKey2007a,koashiSimpleSecurityProof2009,gu2022experimental}. Our protocol adopts a similar single-photon interference approach as twin-field QKD~\cite{lucamariniOvercomingRateDistance2018a,curtySimpleSecurityProof2019,yinFinitekeyAnalysisTwinfield2019}, which allows the key rate to scale with the square root of the total transmittance between the two players. These features simplify implementation by eliminating the need for phase randomization and intensity modulation. By adopting symmetric procedures for all players during key generation and error-rate analysis, our protocol achieves secure key generation without requiring prior knowledge of the dishonest player's identity, thus addressing security loopholes arising from asymmetric basis choices in previous protocols~\cite{shenExperimentalQuantumSecret2023}. We adopt Kato's inequality, which offers tighter bounds compared to the commonly used Azuma's inequality, to provide security against coherent attacks for finite-key analysis~\cite{katoConcentrationInequalityUsing2020,azumaWeightedSumsCertain1967a}. We further optimize the final secure key rates using Kato's inequality~\cite{curras-lorenzoTightFinitekeySecurity2021a,liFinitekeyAnalysisCoherent2024}, achieving transmission distances over 200 km with a misalignment error rate of 1.5\% and exceeding 150 km even with a higher misalignment error rate of 4.5\%. Furthermore, we experimentally demonstrate our QSS protocol under a 30-dB channel loss, including a 5-km fiber distance. We implement nine sets of tests with varying pulse intensities and $X$-basis selection probabilities, analyzing their impacts on secure key rates. We successfully achieve the best key rate of 432 bps with the intensity of $9\times 10^ {-4}$ and the probability of 0.9. \textcolor{black}{While this work focuses on a tripartite QSS protocol, the demand for multiuser quantum communication is growing~\cite{joshi2020trusted,qin2024efficient}. Expanding QSS to accommodate more users is a crucial challenge for future development. Promisingly, our protocol shares similarities with twin-field QKD, which can be extended to multiuser scenarios via time-division multiplexing with a Sagnac network setup~\cite{zhong2022simple}. Exploring similar strategies to increase the user capacity of our protocol will be a key direction for future research.} With its enhanced security, practicality, and high key rates, our protocol contributes to the groundwork for future large-scale quantum multiparty computing networks.

\section{\label{protocol description}PROTOCOL DESCRIPTION}
We show the schematic of our protocol in Fig.~\ref{protocol schematic}. Two symmetric distant participants, Alice and Bob, act as the players of our QSS protocol. They each encode their logic bits and basis selections onto the phase of weak coherent pulses, send them to the central dealer Charlie. Alice and Bob use phase modulators (PMs) to encode the information and variable optical attenuators (VOAs) to dim the pulses. Charlie receives the pulses and, after applying a phase modulation to Bob's pulse, allows them to interfere at the beam splitter. Eventually, Charlie records the detection events of the two superconducting nanowire single-photon detectors, $\rm SPD_1$ and $\rm SPD_2$. The detailed steps of our scheme are as follows:

    \begin{figure}[t]
    \centering
    \includegraphics[width=8.5cm]{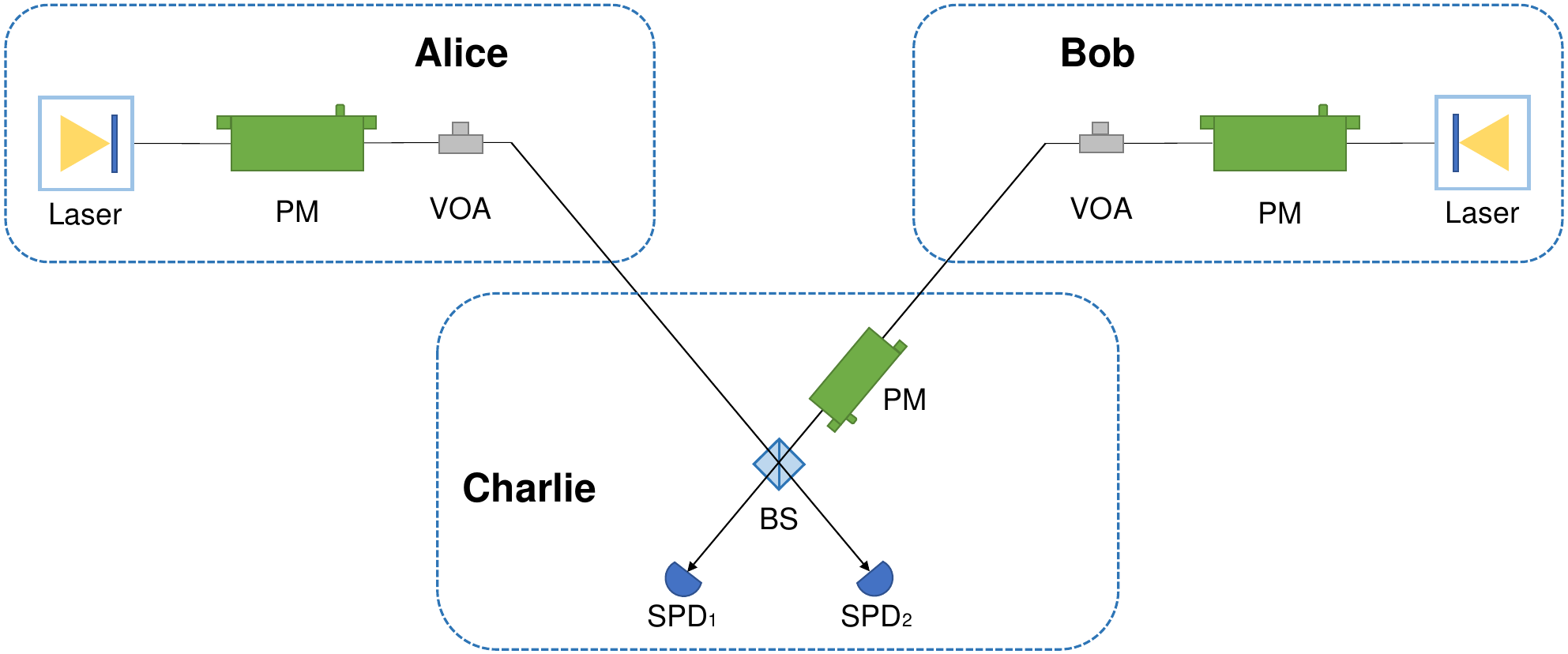}
    \caption{Schematic of our QSS scheme. Alice and Bob send weak coherent pulses to Charlie for interference measurement, with Charlie announcing the corresponding detection results. Alice and Bob encode their pulses with phase modulators (PMs) and use variable optical attenuators (VOAs) to attenuate their pulses to the single-photon level. After adding a phase of \{0, $\pi/2$\} on Bob's pulses based on his basis selection, Charlie performs interference measurement with a beam splitter (BS) and two superconducting nanowire single-photon detectors, $\rm SPD_1$ and $\rm SPD_2$. }
    \label{protocol schematic}
\end{figure}

1. \textbf{Preparation.} In the $i$th round, Alice and Bob each choose a bit value uniformly at random and record them as $s_{a, i}$ and $s_{b, i}$, where $a$ and $b$ denote Alice and Bob, respectively. Then, they each select a basis choice $a_i$ and $b_i$ from $\{X, Y\}$ with probabilities $p_x$ for the $X$ basis and $1-p_x$ for the $Y$ basis. Based on the chosen values, they each prepare a weak coherent pulse to represent their quantum bit and send it to Charlie. If the $X$ basis is chosen, Alice prepares the pulse $\ket{e^{is_{a, i}\pi}\sqrt{\mu}}_A$, and Bob prepares the pulse $\ket{e^{is_{b, i}\pi}\sqrt{\mu}}_B$, where the subscript $A$ and $B$ denote the channel mode of Alice and Bob. If the $Y$ basis is chosen, Alice prepares the pulse $\ket{e^{i(-s_{a, i}+\frac{3}{2})\pi}\sqrt{\mu}}_A$, and Bob prepares the pulse $\ket{e^{i(-s_{b, i}+\frac{3}{2})\pi}\sqrt{\mu}}_B$.

2. \textbf{Measurement.} Charlie selects a basis choice $c_i$ from $\{X, Y\}$ with probabilities $p_x$ for the $X$ basis and $1-p_x$ for the $Y$ basis. Then, he modulates Bob's pulse with an extra phase of $\pi /2$ if the chosen basis is 
$Y$, and leaves Bob's pulse unmodulated if the chosen basis is $X$. After the two pulses from Alice and Bob interfere at a beam splitter (BS), Charlie performs the measurement using two detectors, $\rm SPD_1$ and $\rm SPD_2$. There are four possible outcomes $\{0,1,\emptyset,\perp\}$ where $0$ is the case when $\rm SPD_1$ clicks and $1$ is the case when $\rm SPD_2$ clicks. Cases with no detection are denoted as $\emptyset$. Double detection events are denoted as $\perp$, in which case Charlie randomly chooses a logic bit from $\{0,1\}$. According to the outcomes, Charlie assigns his bit value $s_{c, i}$.

3. \textbf{Basis reconciliation.} After announcing their basis choices over an authenticated public channel, all participants identify the following sets: $\mathcal{X}:=\{i:a_i=b_i=c_i = X \land s_{c, i} \neq \emptyset \}$, $\mathcal{Y}_{bc}:=\{i:b_i=c_i = Y \land a_i = X \land s_{c, i} \neq \emptyset \}$, and $\mathcal{Y}_{ac}:=\{i:a_i=c_i = Y \land b_i = X \land s_{c, i} \neq \emptyset \}$, where the subscripts $a$, $b$, and $c$ denote Alice, Bob, and Charlie, respectively. They then verify that the sizes of these sets meet the required thresholds for key generation: $|\mathcal{X}| \geq n_{X}$, $|\mathcal{Y}_{bc}| \geq n_{Y_{bc}}$, and $|\mathcal{Y}_{ac}| \geq n_{Y_{ac}}$, where $n_{X}$, $n_{Y_{bc}}$, and $n_{Y_{ac}}$ are predetermined thresholds correlated with the desired key length. Steps 1-3 are repeated until these conditions are met, and the number of rounds required to fulfill these conditions is denoted as $N$.

4. \textbf{Raw key generation and error estimation.} Based on the basis reconciliation results, the participants sift their bits. The bit pairs satisfying $S_X := \{\{s_{a, i},s_{b, i}, s_{c, i}\}: i \in \mathcal{X}\}$ are used to generate the secure key bits, with a portion of these bits dedicated to analyzing the bit error rate $E_b^X$. 
For the bit pairs satisfying $\{s_{a, i},s_{b, i}, s_{c, i}\}: i \in \mathcal{Y}_{ac}$, Charlie's corresponding bit is flipped: $s_{c, i}^\prime = s_{c, i} \oplus 1$. This forms a set of key bits $S_{Y_{ac}} := \{\{s_{a, i},s_{b, i}, s_{c, i}^\prime\}: i \in \mathcal{Y}_{ac}\}$. These bits, along with those satisfying $S_{Y_{bc}} := \{\{s_{a, i},s_{b, i}, s_{c, i}\}: i \in \mathcal{Y}_{bc}\}$, are revealed and used independently to calculate the phase error rates $E_{p_{ac}}$ and $E_{p_{bc}}$. The higher of these two error rates is then used to calculate the upper bound of the final phase error rate $\overline{E}_{p}$. Finally, the error rates are checked against predetermined thresholds, $E_b^{\rm tol}$ for the bit error rate and $E_p^{\rm tol}$ for the phase error rate. If either $E_b^X > E_b^{\rm tol}$ or $\overline{E}_{p}> E_p^{\rm tol}$, indicating potential information leakage exceeding security requirements, the protocol aborts. Otherwise, the protocol proceeds to the next step.

5. \textbf{Postprocessing.} All participants perform error correction, where the amount of information leakage, $\lambda_{\rm{EC}}$, is revealed. Then, to ensure that they share data with the perfect correlation, they perform an error-verification step using a random universal$_{2}$ hash function that publishes $\log_{2}2/\epsilon_c$ bits of information.
Following successful error correction, privacy amplification is performed to distill the final secure keys: $S_a$, $S_b$, and $S_c$ by applying a random universal$_{2}$ hash function.  The final keys satisfy the following conditions: $S_c = S_a \oplus S_b$ and $|S_a|=|S_b|=|S_c| = \ell$ bits, \textcolor{black}{in which $\ell$ represents the length of the final generated key}.

\section{\label{security analysis}SECURITY ANALYSIS}
This section provides a comprehensive security analysis of our proposed protocol. We assume a perfect source is utilized in our protocol to simplify the analysis. We analyze the protocol step-by-step to ensure the generated secure key meets the requirements of a secret sharing scheme. We then demonstrate our protocol's resistance against malicious players attempting to steal information from other participants. We focus solely on internal eavesdroppers, as the potential information leakage to them exceeds that of external eavesdroppers, ensuring this approach does not compromise the protocol's overall security. We establish an equivalence between the security of our protocol against internal eavesdroppers and that of phase-encoded QKD. This allows us to leverage the well-established security of phase-encoded QKD to effectively verify the security of our QSS protocol~\cite{loSecurityQuantumKey2007a,koashiSimpleSecurityProof2009}. Subsequently, we show that the security analysis of the protocol is symmetrical concerning the two different players, avoiding the security loopholes introduced by asymmetric protocol steps. Finally, we obtain the secure key rate of our protocol by analyzing the information leakage that a malicious player can obtain from other participants.

\subsection{Secure key bit correlation}
In the $i$th round, Alice and Bob each randomly choose a bit value, recorded as $s_{a, i}$ and $s_{b, i}$, respectively. Charlie then sets his bit value, $s_{c, i}$, based on the measurement outcome. QSS protocols require a specific correlation between the key bits. This involves dividing a secret from the dealer into two parts and distributing them to two players in such a way that both parts of the split secret are necessary to reconstruct the original secret. This is achieved by setting Charlie's raw key bits to the exclusive \textsmaller{OR (XOR)} of Alice and Bob's raw key bits, denoted as $S_c = S_a \oplus S_b$. We will show that the keys generated by our protocol satisfy this correlation.
\par As shown in the first step of our protocol, Alice (Bob) prepares weak coherent states denoted as $\ket{e^{is_{a, i}\pi}\sqrt{\mu}}_A$ ($\ket{e^{is_{b, i}\pi}\sqrt{\mu}}_B$) under the $X$ basis, and $\ket{e^{i(-s_{a, i}+\frac{3}{2})\pi}\sqrt{\mu}}_A$ ($\ket{e^{i(-s_{b, i}+\frac{3}{2})\pi}\sqrt{\mu}}_B$) under the $Y$ basis, with $\mu$ denoting the pulse intensity. Due to the $2\pi$ periodicity, the weak coherent states under the $Y$ basis can also be expressed as $\ket{e^{i(s_{a, i}-\frac{1}{2})\pi}\sqrt{\mu}}_A$ ($\ket{e^{i(s_{b, i}-\frac{1}{2})\pi}\sqrt{\mu}}_B$). In the second step, Charlie modulates an extra $0$ ($\pi/2$) phase to Bob's pulse under $X$ ($Y$) basis.
\par The two optical modes sent by Alice and Bob can be described using the annihilation operators $a_A$ and $a_B$, respectively: $\ket{\alpha}_A = e^{-|\alpha|^2/2}e^{\alpha a_A^\dagger}\ket{0}_A$ and $\ket{\alpha}_B = e^{-|\alpha|^2/2}e^{\alpha a_B^\dagger}\ket{0}_B$, where $\alpha = \sqrt{\mu}e^{i\theta}$ represents the complex amplitude of the coherent state and $\ket{0}$ represents the vacuum state, with $\mu$ being the laser intensity and $e^{i\theta}$ the laser phase. The two modes are combined in the beam splitter after their phases are modulated by the phase modulators. Based on the quantum description of the beam splitter, the outgoing annihilation operators $a_1$ and $a_2$, corresponding to the two single-photon detectors $\rm SPD_1$ and $\rm SPD_2$, respectively, are expressed as $a_1 = (a_B + a_A)/\sqrt{2}$ and $a_2 = (a_B - a_A)/\sqrt{2}$. Since the two pulses have the same intensity $\mu$, the response of $\rm SPD_1$ or $\rm SPD_2$ is determined solely by the phase difference between the two pulses incident on the beam splitter. If the phase difference is 0, only SPD1 will click. If the phase difference is $\pi$, only SPD2 will click. If the phase difference is $\pi/2$ or $3\pi/2$, both detectors have an equal probability of clicking. For the key bits in $S_X$ and $S_{Y_{bc}}$ generated during the $i$th round, the phase difference is given by $\Delta \Phi =  (s_{b, i} - s_{a, i})\pi$. Charlie sets his logic bit to 0 (1) upon $\rm SPD_1$ ($\rm SPD_2$) clicks, resulting in $s_{c, i} = s_{a, i} \oplus s_{b, i}$. For the key bits in $S_{Y_{ac}}$ generated during the $i$th round, the phase difference is given by $\Delta \Phi =  (s_{b, i} - s_{a, i} + 1)\pi$. With Charlie's bit flipped: $s_{c, i}^\prime = s_{c, i} \oplus 1$, the key bit correlation remains $s_{c, i}^{\prime} = s_{a, i} \oplus s_{b, i}$. Thus, the final keys $S_a$, $S_b$, and $S_c$ satisfy the correlation: $S_c = S_a \oplus S_b$.
\par It is worthwhile to note that our discussion thus far assumes ideal scenarios. Under these ideal conditions, the generated secure key fulfills the requirements of a secret sharing scheme.

\subsection{Security equivalence}

\begin{figure}[b]
    \centering
    \includegraphics[width=8.5cm]{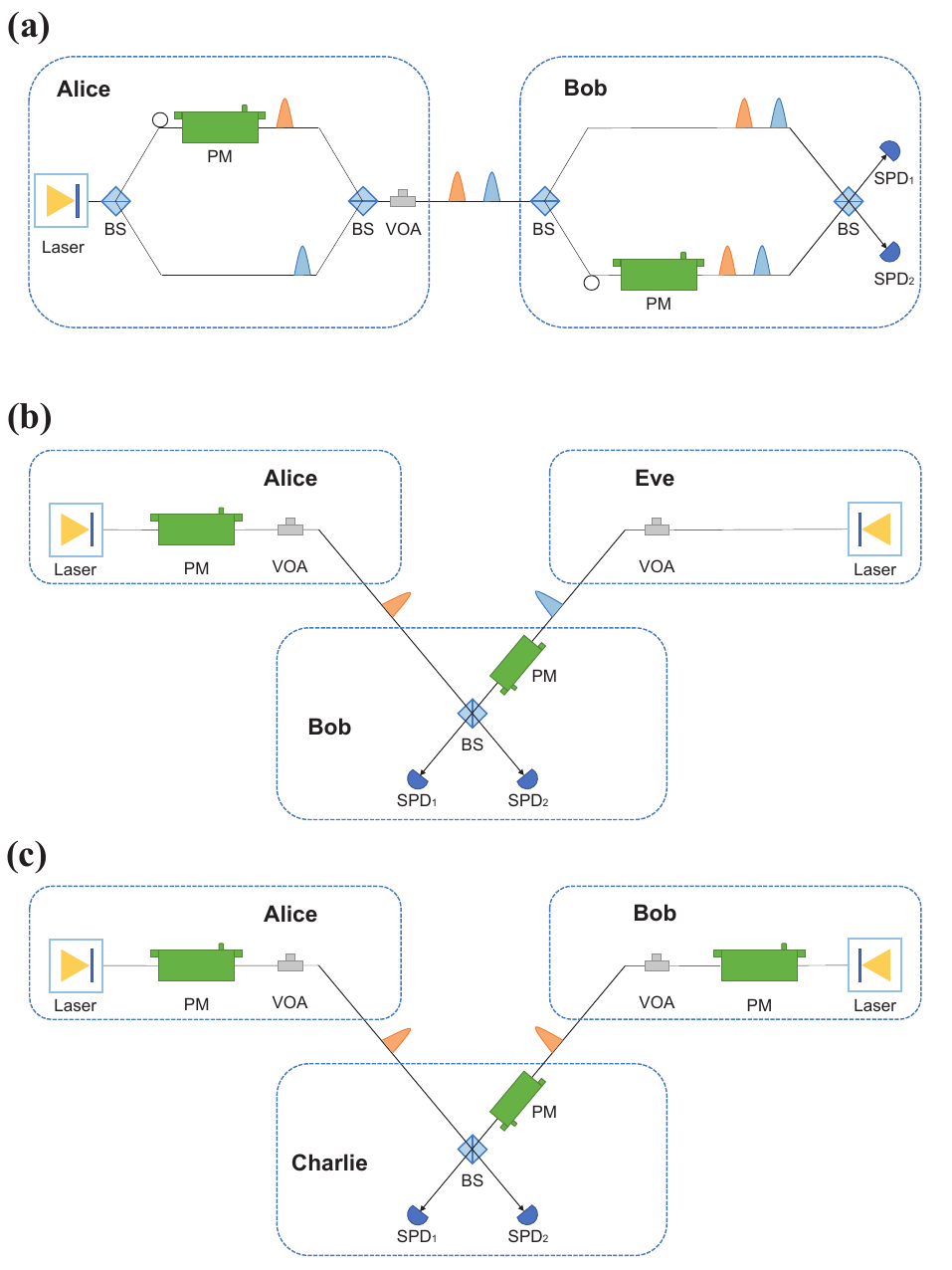}
    \caption{Schematic diagrams of the phase-encoded QKD protocol, its variant, and our QSS protocol. \textbf{(a)} Schematic of the original phase-encoded QKD protocol. Each round, Alice prepares a signal pulse (orange) using her phase modulator and a reference pulse (blue) before sending them to Bob. Bob modulates the phases of the pulses and then performs a two-pulse interference. The phase difference is obtained to generate a secure key bit. \textbf{(b)} An equivalent variant of the phase-encoded protocol, where an eavesdropper Eve sends the reference pulse instead of Alice. \textbf{(c)} Schematic of our QSS protocol, where Bob replaces the eavesdropper and modulates his pulse using a phase modulator.}
    \label{equivalent protocols}
\end{figure}

We provide a detailed description of the security equivalence between our QSS protocol and the phase-encoded QKD protocol. To establish the security equivalence, we first introduce the phase-encoded QKD protocol, as shown in Fig.~\ref{equivalent protocols} \textbf{(a)}. The phase-encoded QKD protocol involves two participants, Alice and Bob. Alice chooses a basis out of $X$ and $Y$. Based on the chosen basis, Alice emits a signal pulse and encodes her logic bit on the phase of the signal pulse. In $X$ basis, Alice emits a weak coherent pulse $\ket{e^{is_{a, i}\pi}\sqrt{\mu}}_A$, where $s_{a, i}$ is the logic bit of Alice in the $i$th round and $\mu$ is the intensity of the pulse. In $Y$ basis, Alice emits a weak coherent pulse $\ket{e^{i(-s_{a, i}+\frac{3}{2})\pi}\sqrt{\mu}}_A$. Alice also emits a reference pulse $\ket{\sqrt{\mu}}_A$ as an ancillary pulse and sends both prepared pulses to Bob. Bob chooses a basis out of $X$ and $Y$ and applies a phase modulation of 0 ($\pi$) on the signal pulse when the chosen basis is $X$ ($Y$). Bob then allows the two pulses to interfere and measures the phase difference, obtaining his logic bit based on the measurement results. Alice and Bob retain their logic bits as raw key bits when they choose the same basis and discard them otherwise. In the $X$ basis, they use the raw key bits to generate the final key and estimate the bit error rate. In the $Y$ basis, they use the raw key bits to bound the phase error rate. 

\par We then consider a variant of our scheme and demonstrate the equivalence between its security against an internal eavesdropper and the security of the phase-encoded QKD protocol. As an intermediate step, we consider a variant of the phase-encoded QKD protocol, as shown in Fig.~\ref{equivalent protocols} \textbf{(b)}. In this variant, Alice transmits only the signal pulses, while an eavesdropper, Eve, sends the reference pulses. These pulses arrive simultaneously at Bob's location and interfere. In the phase-encoded QKD protocol, even if an eavesdropper knows the relative phase of the reference pulse, they cannot obtain information about the secret key~\cite{loSecurityQuantumKey2007a}. This ensures the security of the variant protocol. 
\par Finally, regardless of whether Alice or Bob acts as the eavesdropper, our QSS protocol is equivalent to the variant of the phase-encoded QKD protocol. As shown in Fig.~\ref{equivalent protocols} \textbf{(c)},  this variant retains the preparation and measurement steps of the original scheme but modifies the sifting and parameter estimation steps. Since the potential phase information added by an eavesdropper does not compromise the security of phase-encoded QKD, and our protocol demonstrates equivalent security against internal eavesdroppers, we have established the security of our QSS protocol against an internal malicious player.
 
\subsection{Eavesdropper's identity}
\par We establish the security of our protocol by demonstrating its equivalence to the phase-encoded QKD protocol, thus leveraging existing security proofs~\cite{loSecurityQuantumKey2007a,koashiSimpleSecurityProof2009}. However, existing security proofs for the phase-encoded QKD protocol assume honest participants with complete trust, while eavesdroppers are completely distrusted. In our protocol, the malicious player's identity is revealed only during the use of raw key bits to bound error rates and when the two players collaborate to obtain information from Charlie. Before this stage, we have no information about the malicious player's identity. Therefore, the basis choice in the sifting process must be symmetric for both Alice and Bob during the generation of secure key bits. This symmetry is crucial to ensure protocol security even if Alice and Bob's roles were interchanged. As demonstrated in our protocol the key bits in $S_X$ are used to form their secure key and use a portion of it to estimate the bit error rate. This process is symmetric for both Alice and Bob. Ideally, they would disclose the logic bits under the corresponding chosen bases to bound the phase error rate based on the internal eavesdropper's identity. Specifically, if Alice were the internal eavesdropper, the logic bits in $S_{Y_{bc}}$ would be revealed to bound the phase error rate; if Bob were the internal eavesdropper, the logic bits in $S_{Y_{ac}}$ would be revealed instead. However, since the internal eavesdropper's identity remains unknown, both steps must be executed in the protocol's practical implementation. The higher of the two resulting values, representing the worst-case information leakage scenario, is selected as the phase error rate.

\subsection{Information leakage}
We now analyze the security of the variant scheme in the scenario where Bob acts as the internal eavesdropper. Since Bob acts as the eavesdropper, his secure key information is naturally fully known to the eavesdropper. Therefore, our security analysis of the variant focuses on the potential leakage of Alice's pulse phase information. This analysis is conducted by introducing a ``quantum coin"  into our system and considering an equivalent protocol. 
\par \textcolor{black}{Imagine a coin that Alice flips for each signal she sends. The coin has two sides, representing her two choices of encoding basis for the key bit. If the coin is perfectly fair (50/50 probability for each side), Eve gains no information about Alice's basis choice just by observing the coin flips. This corresponds to the ideal scenario where the signals themselves reveal nothing about the chosen basis.}
\par \textcolor{black}{However, in a practical scenario with weak coherent states, the signals are not perfectly basis independent. This is analogous to Alice having a slightly biased coin. The bias, though small, could allow Eve to gain partial information about the chosen basis, potentially compromising the protocol's security.}
\par \textcolor{black}{The ``quantum coin" in our analysis precisely captures this basis dependence. It is not a physical coin, but a mathematical tool to quantify how much information Eve could extract about Alice's basis choice based on the inherent properties of the emitted signals.}
\par We first denote Alice's basis-dependent states as
\begin{align}
    &\ket{\Psi_x}=(\ket{0_X}\otimes \ket{\alpha}+\ket{1_X}\otimes \ket{-\alpha})/\sqrt{2} \\
    &\ket{\Psi_y}=(\ket{1_Y}\otimes \ket{i\alpha}+\ket{0_Y}\otimes \ket{-i\alpha})/\sqrt{2} 
\end{align}
where $\ket{0_X}$, $\ket{1_X}$ are eigenstates of the Pauli operator $\sigma_x$ and $\ket{0_Y}$, $\ket{1_Y}$ are eigenstates of the Pauli operator $\sigma_y$. Then, Alice measures the state of the quantum coin under $\ket{0_Z}$, $\ket{1_Z}$ basis and chooses her states $\ket{\Psi_x}$ or $\ket{\Psi_y}$ according to the result. Note that we assume the measurement is delayed till the eavesdropper finishes eavesdropping on the signals. We can denote the joint state of the quantum coin state and Alice's source state as 
\begin{equation}
    \ket{\Phi}=\sqrt{p_x}\ket{0_Z}\otimes\ket{\Psi_x}+\sqrt{p_y}\ket{1_Z}\otimes\ket{\Psi_y},
\end{equation}
where $p_x$ and $p_y$ denote the probabilities of Alice selecting the $X$ basis and $Y$ basis, respectively, to encode her signals. Introducing the quantum coin establishes a relationship between the basis dependence of Alice's signal and the imbalance $\Delta$ of the quantum coin. This allows us to quantify the basis dependence and bound the information leakage of Alice's pulses.
\par The logic bits in $S_{Y_{ac}}$ and $S_{Y_{bc}}$ are used to calculate the phase errors. The basis dependence of Alice's signals, denoted by $\Delta$, is calculated as
\begin{equation}
    1-2Q_{\mu}\Delta = \braket{\Psi_y|\Psi_x},
\end{equation}
where $Q_{\mu}$ denotes the protocol's total gain. The phase error rate $E_p$ is then calculated using the formula
\begin{align} \label{Yb2Xp}
    E_p = &E_b^Y + 4\Delta(1-\Delta)(1-2E_b^Y)\\
    & + 4(1-2\Delta)\sqrt{\Delta(1-\Delta)E_b^Y(1-E_b^Y)},
\end{align}
where $E_b^Y$ is the bit error rate calculated using the key bits in $S_{Y_{ac}}$ or $S_{Y_{bc}}$.
\par The final key rate of our protocol is expressed as
\begin{equation}
    R = Q_{\mu}[1-f_eH(E_b^X)-H(E_p)],
\end{equation}
where $f_e$ is the error-correction efficiency, $E_b^X$ is the bit error rate of the raw key bits in $S_X$, and $H(x)=-x\log_2 x -(1-x)\log_2(1-x)$ is the Shannon entropy. The total gain $Q_{\mu}$ for the key bits in $S_X$ is calculated as $Q_{\mu}=(1-p_d)[1-(1-2p_d)e^{-2\mu\eta}]$, where $p_d$ is the dark count rate of both of Charlie's detectors and $\mu$ is the light intensity of the weak coherent states. The bit error rate $E_b^X$ is given by $E_b^X Q_{\mu}=e_d(1-p_d)[1-(1-p_d)e^{-2\mu\eta}]+(1-e_d)p_d(1-p_d)e^{-2\mu\eta}$, where $e_d$ is the misalignment error rate of the detectors.

\section{\label{simulations}SIMULATIONS}

This section evaluates the performance of our protocol in the finite-key regime and practical implementations through theoretical simulations. We first clarify the specific security criteria employed in our analysis. For small errors in our protocols $\epsilon_c, \epsilon_s > 0$, our protocol is $\epsilon_c +\epsilon_s$ secure if it is $\epsilon_c$ correct and $\epsilon_s$ secret. The $\epsilon_c$-correctness condition is satisfied if $\Pr [S_c \neq S_a \oplus S_b] \leq \epsilon_c$, i.e., the secret key bits satisfy the correlation except with a small probability $\epsilon_c$. The $\epsilon_s$-secrecy condition is satisfied if $(1-p_{\rm abort})||\rho_{CE}-U_C\otimes \rho_E||_1/2 \leq \epsilon_s$ where $\rho_{CE}$ is the classical-quantum state describing the joint state of $S_c$ and the system of the eavesdropper $E$, $U_C$ is the uniform mixture of all possible values of $S_c$, and $p_{\rm abort}$ is the probability that the protocol aborts. The symbol $\epsilon_{PA}$ denotes the failure probabilities of privacy amplification. In practice, we have $\epsilon_s = \sqrt{\epsilon} + \epsilon_{PA}$, where $\epsilon$ is the failure probability of the phase error rate estimation.

\begin{figure}[b]
    \centering
    \includegraphics[width=8.5cm]{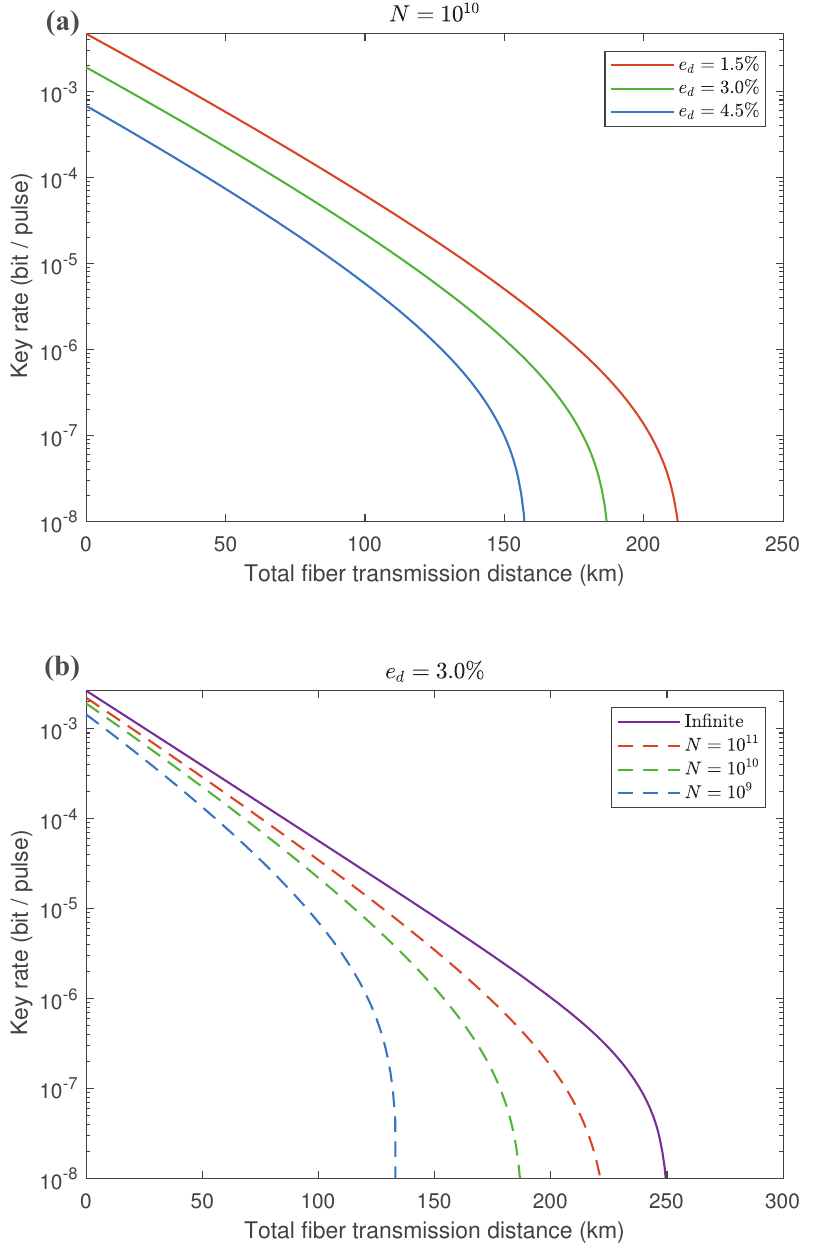}
    \caption{Secure key rates of our protocol under the finite-key regime for different misalignment error rates $e_d$ and total numbers of pulses $N$. \textbf{(a)} The total number of pulses $N$ is fixed at $10^{10}$, a typical value in practical deployments. The misalignment error rates $e_d$ are set to $1.5\%$, $3.0\%$, and $4.5\%$. The key rates demonstrate that our protocol can generate secure keys over distances exceeding 200 km with a relatively low misalignment error rate of $1.5\%$ in practical implementations. \textbf{(b)} The misalignment error rate $e_d$ is fixed at $3.0\%$, while the values of the total number of pulses are $N = 10^9$, $10^{10}$, $10^{11}$, and an infinite regime. As $N$ increases, the secure key rates of our protocol approach the performance of the ideal infinite-key scenario, with distances exceeding 200 km achieved when $N = 10^{11}$.}
    \label{sim results}
\end{figure}

\par In the finite-key regime, conditioned on passing the error-estimation and error-verification steps, the secure key rate of our protocol is given by
\begin{align}
    \ell=&n_X[1-H(\overline{E}_{p})]-\lambda_{\rm{EC}}-\log_2\frac{2}{\epsilon_c}-\log_2\frac{1}{4\epsilon^2_{PA}},
\end{align}
where $n_X$ denotes the size of the set $\mathcal{X}$. The term $\lambda_{\rm{EC}}$ $ = n_{X}f_eH(E_b^X)$ denotes the number of bits consumed during the error-correction process. Here, $f_e$ denotes the error-correction efficiency, and $E_b^X$ denotes the bit error rate of the key bits in $S_X$. A detailed analysis of the upper bound for the observed phase error rate $\overline{E}_{p}$ is provided in Appendix~\ref{appendix A}.

\par We consider the distance between Alice and Bob as $L$, resulting in their distances to the central participant Charlie being $L/2$ each. This leads to a total channel transmittance $\eta$ of $\eta_d \times 10^{-\alpha L/ 20}$, where $\eta_d$ is the detection efficiency of Charlie's detectors and $\alpha$ is the attenuation coefficient of the ultra-low-loss fiber used in our simulations. This demonstrates that the secure key rate of our protocol scales with the square root of the total distance between Alice and Bob, similar to the twin-field quantum key distribution setup~\cite{lucamariniOvercomingRateDistance2018a}. Our simulations use parameters conforming to the actual experimental apparatus: $\eta_d = 40\%$, $p_d = 2\times 10^{-8}$, $\alpha = 0.167$, and $f_e = 1.16$. Under these conditions, we incorporate pulse intensities and $X$-basis selection probabilities as parameters into the optimization algorithm to obtain optimal secure key rates for different distance scenarios. As shown in Fig.~\ref{sim results} \textbf{(a)}, we fix the total number of pulses to $N=10^{10}$, a typical value in real experimental deployments. By setting different misalignment error rates $e_d$ to $1.5\%$, $3.0\%$, and $4.5\%$, we obtain the secure key rates of our protocol under different misalignment error rates. The simulation result from Fig.~\ref{sim results} \textbf{(a)} shows that our protocol can achieve a transmission distance exceeding 150 km under the condition of the misalignment error rates $e_d = 4.5\%$. This suggests the practicality of our protocol in field tests and real-world implementations, where misalignment errors are likely to occur. Figure.~\ref{sim results} \textbf{(b)} presents the simulation results for a fixed misalignment error rate $e_d = 3.0\%$ and varying total numbers of pulses $N = 10^9$, $10^{10}$, $10^{11}$, and infinite. As shown, the secure key rates of our protocol increase with the total number of pulses, achieving transmission distances exceeding 200 km when $N = 10^{11}$. A detailed finite-key analysis is provided in Appendix~\ref{appendix A}.

\section{\label{experiment}EXPERIMENTAL DEMONSTRATION}

\begin{figure*}[t]
    \centering
    \includegraphics[width=17cm]{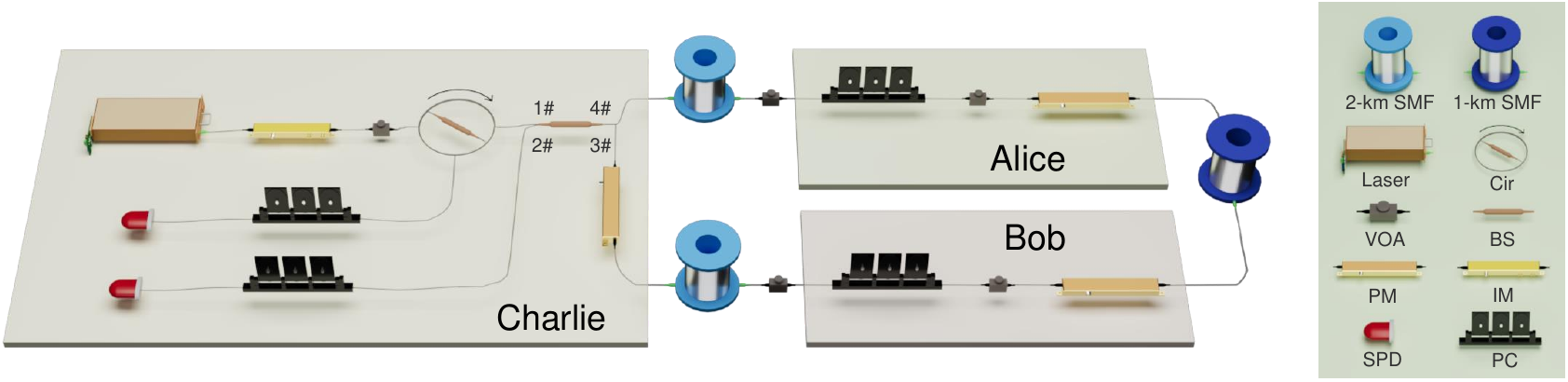}
    \caption{Schematic of our QSS experimental setup. Continuous wave is chopped by two intensity modulators (IMs) into a pulse train, which is split into two identical pulses by a 50:50 BS. The two pulses enter the Sagnac loop and are modulated by Alice, Bob, and Charlie, respectively. After traveling through the loop, the two modulated pulses interfere at Charlie's BS and are detected by two SPDs. SMF, single-mode fiber; Cir, circulator; PC, polarization controller.}
    \label{experimental setup}
\end{figure*}

Our QSS experimental setup is depicted in Fig.~\ref{experimental setup}. On Charlie's side, a continuous-wave laser source (NKT Koheras BASIK E15) with a center wavelength of 1550.12 nm and a linewidth less than 0.1 kHz is used. Two intensity modulators (IMs) are used to chop the continuous wave into a pulse train, achieving a pulse extinction ratio of over 30 dB. The resulting pulses have a temporal width of less than 1 ns and a repetition rate of 100 MHz. Subsequently, the pulses pass through a circulator (Cir) and are then split into two identical pulses by a 50:50 BS.
\par The two pulses enter the Sagnac loop, which guarantees phase stability between Alice and Bob. Within the loop, the pulses are modulated by Alice, Bob, and Charlie. Notably, Alice (Bob) uses a PM to encode the counterclockwise (clockwise) pulses exclusively. The basis selection is performed with probabilities $p_x$ and $p_y$ for the $X$ and $Y$ bases. A random phase of 0 or $\pi$ is added for the $X$ basis, while a phase of $\pi/2$ or $3\pi/2$ is added for the $Y$ basis. Specifically, Charlie randomly adds a phase of 0 or $\pi/2$ to Bob's pulses with probabilities $p_x$ and $p_y$, respectively. The system includes 2-km single-mode fibers (SMFs) connecting Alice and Bob to Charlie, and a 1-km SMF connecting Alice and Bob. Additionally, polarization controllers (PCs) are placed between the PMs and SMFs to ensure polarization alignment.

\begin{figure}[b]
    \centering
    \includegraphics[width=8.5cm]{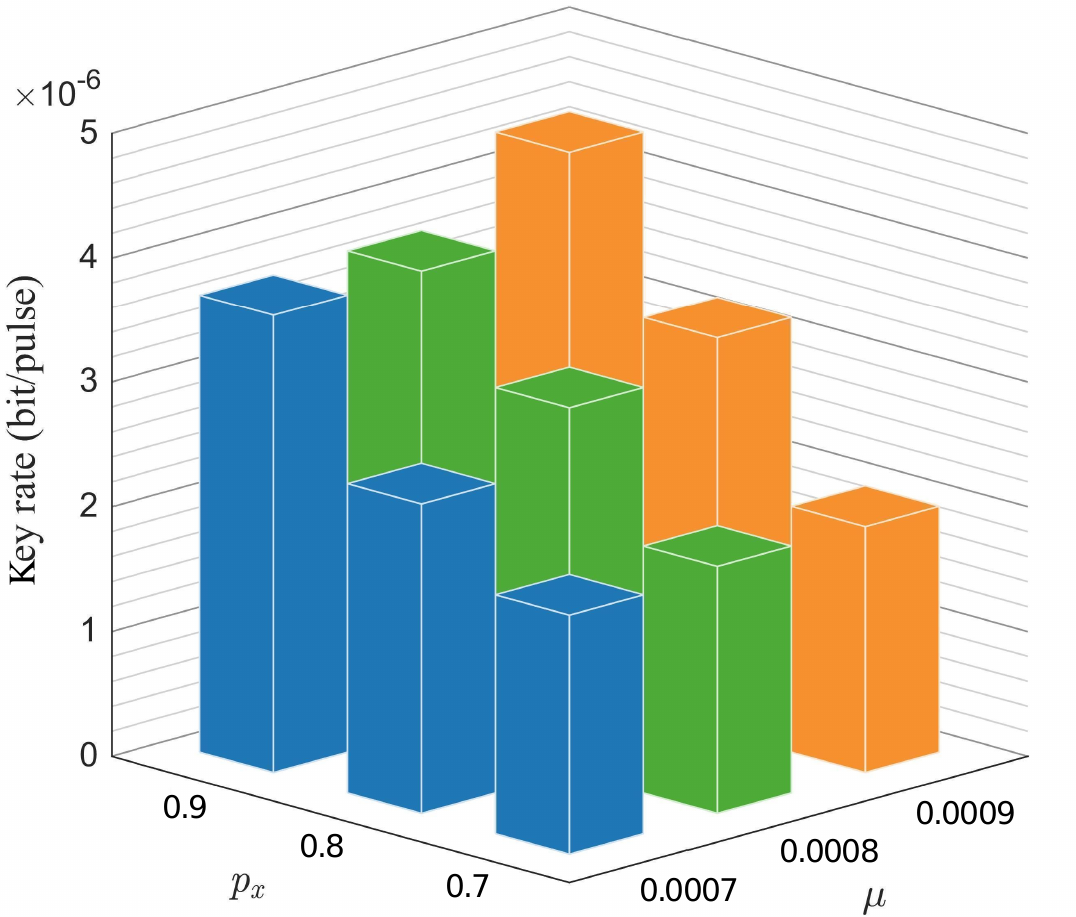}
    \caption{Key rates of our QSS scheme vs pulse intensities, $\mu$, and $X$-basis selection probabilities, $p_{x}$. Varying the intensities $7\times 10^ {-4}$, $8\times 10^ {-4}$ and $9\times 10^ {-4}$and $X$-basis selection probabilities 0.9, 0.8, and 0.7, yields secure key rates ranging from 432 to 192 bps.}
    \label{experimental result}
\end{figure}

\par The two modulated pulses interfere at Charlie's BS. Charlie uses two SPDs to detect the interference results. The detection efficiencies of SPDs are 39.3\% and 39.5\%, with dark count rates of $2.8\times 10^{-8}$ and $4.0\times 10^{-8}$ within 2-ns time windows. To compensate for the differing detection efficiencies, which affect the ratio of 0 and 1 in the raw keys, Charlie employs a PM to add an extra phase of 0 or $\pi$ to Bob's pulses with equal probability (50\% each). The total channel attenuation between Alice and Bob is 30 dB. To simulate a symmetric channel, an additional attenuation of approximately 2.5 dB is added between Alice and Charlie to balance Charlie's PM loss.
\par We implemented our protocol and investigated the impact of the pulse intensity $\mu$ and the selection probability of $X$ basis $p_{x}$ on secure key rates. Theoretical simulation predicts the highest key rate at $\mu = 9\times 10^ {-4}$ when $p_{x}$ is 0.9. To optimize performance, we, respectively, chose the intensities of $9\times 10^ {-4}$, $8\times 10^ {-4}$, and $7\times 10^ {-4}$ and the probabilities of 0.9, 0.8, and 0.7 to perform nine sets of tests. The accumulation time is 500 seconds and the data size for each test is $5\times 10^ {10}$. Detailed experimental data under different input values of $\mu$ and $p_{x}$ are provided in Appendix~\ref{appendix B}. After collecting all data, we calculated the experimental quantum bit error rates (QBERs), phase error rates, and secure key rates. Table~\ref{experimental data} lists the experimental data. 
\par As shown in Fig.~\ref{experimental result}, the highest key rate 432 bps is achieved with $p_x = 0.9$ and $\mu = 9\times 10^{-4}$. With the decrease of $p_x$ and $\mu$, the final key rate tends to be lower. Since the detection events under $X$ basis are used to generate final key bits, the key rates is directly related to $p_{x}$. The results confirm that reducing the $X$-basis selection probability leads to a decrease in the final key rate under the same pulse intensity. With higher intensity, the number of detections increases, but more information is leaked, leading to an increased phase error rate.
Additionally, other experimental parameters, such as QBERs in the $X$ basis and $Y$ basis, also influence the results. For a fixed $X$-basis selection probability, both pulse intensity and error rate affect the key rate. Therefore, as the intensity decreases, the key rate under the same probability of $X$ basis does not always decrease. The largest QBER observed in our experiment is less than 1.8\%, which shows our protocol can resist environmental interference and has robustness.

\section{\label{conclusion}CONCLUSION}

\begin{table*}[t]
    \renewcommand{\thetable}{I}
    \centering
    \caption{Summary of experimental results. The total data size is $5\times 10^ {10}$. $p_{x}$ denotes the $X$-basis selection probability, $\mu$ is the higher intensity of the pulses sent by Alice and Bob, $E_b^X$ and $E_b^Y$ are the experimental quantum bit error rates in the $X$ and $Y$ bases, respectively, $E_p$ is the phase error rate, $n_{x}$ and $n_{y}$ are the numbers of detections in the $X$ and $Y$ bases, respectively, and SKR is the secure key rate.}
    \label{experimental data} 
    \setlength{\tabcolsep}{16.5pt}
    \begin{tabular*}{\textwidth}[c]{cccccccc}
        \toprule
	$p_{x}$ & {$\mu$} & {$E_b^X(\%)$} & {$E_b^Y(\%)$} & $E_p(\%)$ & {$n_{x}$} & {$n_{y}$} & {$\rm SKR (bit/pulse)$} \\
	\midrule
        0.9 & $9\times 10^ {-4}$ & 0.95 & 1.16 & 15.66 & 787 407 & 18 056 & $4.32\times 10^ {-6}$\\
	      & $8\times 10^ {-4}$ & 1.03 & 1.37 & 15.55 & 683 629 & 15 889 & $3.70\times 10^ {-6}$\\
	    & $7\times 10^ {-4}$ & 1.05 & 1.30 & 14.18 & 606 878 & 13 769 & $3.67\times 10^ {-6}$\\
 	0.8 & $9\times 10^ {-4}$ & 1.05 & 1.27 & 14.99 & 561 372 & 72 686 & $3.16\times 10^ {-6}$\\
	    & $8\times 10^ {-4}$ & 0.99 & 1.44 & 14.57 & 494 329 & 63 721 & $2.92\times 10^ {-6}$\\
	    & $7\times 10^ {-4}$ & 1.12 & 1.79 & 14.46 & 430 832 & 55 912 & $2.48\times 10^ {-6}$\\
	0.7 & $9\times 10^ {-4}$ & 1.07 & 1.57 & 15.66 & 377 194 & 133 361 & $1.97\times 10^ {-6}$\\
	    & $8\times 10^ {-4}$ & 0.99 & 1.44 & 14.32 & 330 782 & 117 229 & $1.98\times 10^ {-6}$\\
	    & $7\times 10^ {-4}$ & 0.92 & 1.53 & 13.37 & 292 026 & 104 446 & $1.92\times 10^ {-6}$\\
	\bottomrule
    \end{tabular*}
\end{table*}

In summary, we have proposed a QSS protocol utilizing phase encoding on weak coherent states to eliminate the need for phase randomization and intensity modulation. The same single-photon interference method as the \textcolor{black}{twin-field} QKD is used in our protocol, allowing the key rate to scale with the square root of the channel transmittance. By employing a symmetric basis choice for all participants during secure key generation and error-rate analysis, our protocol achieves secure key generation without prior knowledge of the dishonest player's identity, addressing the security loopholes associated with asymmetric basis choices compared to the previous protocols. Additionally, we employed Kato's inequality instead of the more commonly used Azuma's inequality to ensure security against coherent attacks for the finite-key analysis. Following established practices, we addressed the optimization procedures of Kato's inequality to further improve the final secure key rate. In simulations, our protocol achieves transmission distances exceeding 200 km with a misalignment error rate of 1.5\%, and over 150 km even with a higher misalignment error rate of 4.5\%, demonstrating its applicability in field tests.
\par Furthermore, we experimentally demonstrate our QSS protocol using nine sets of tests at a 30-dB channel loss, including a 5-km fiber distance. Varying the pulse intensities and $X$-basis selection probabilities, we achieve secure key rates ranging from 432 to 192 bps. In our implementation, the key factor influencing the final key rates is the $X$-basis selection probability. A high proportion of $X$ basis results in a smaller number of detection events under the $Y$ basis, implying increasing statistical fluctuation of $Y$ basis and ultimately decreasing the secure key rate. When the dataset is sufficiently large to neglect statistical fluctuations, increasing the selection probability can further improve the secure key rate. Additionally, pulse intensity and QBERs in the $X$ and $Y$ bases also influence the key rate. Increasing the pulse intensity can enhance key rates, but at the risk of leaking more information and increasing the phase error rate. Therefore, choosing appropriate pulse intensities and $X$-basis selection probabilities is crucial for optimizing the final key rates.
\par \textcolor{black}{Looking ahead, the development of multiuser QSS protocols is essential for realizing large-scale quantum networks and enabling more complex multiparty quantum communication tasks. Drawing inspiration from multiuser twin-field QKD implementations~\cite{zhong2022simple}, we plan to investigate strategies for extending our QSS protocol to a greater number of users. This will involve exploring techniques like time-division multiplexing, advanced synchronization methods, and potentially alternative network architectures to address the challenges of scalability in QSS.}
Having achieved information-theoretic security and validation in real optical systems, our protocol offers a valuable reference for practical applications of multiparty quantum communication and the realization of quantum multiparty computing networks.

\begin{acknowledgments}
This work is supported by the National Natural Science Foundation of China (Grant No. 12274223), the Fundamental Research Funds for the Central Universities and the Research Funds of Renmin University of China (Grant No. 24XNKJ14), and the Program for Innovative Talents and Entrepreneurs in Jiangsu (Grant No. JSSCRC2021484).
\end{acknowledgments}

\appendix
\section{\label{appendix A}FINITE-KEY ANALYSIS}
In this work, we use Kato's inequality instead of the commonly used Azuma's inequality as concentration inequality to provide security against coherent attacks by calculating the upper bound of the observed phase error rate $\overline{E}_{p}$, taking statistical fluctuations into account~\cite{katoConcentrationInequalityUsing2020,azumaWeightedSumsCertain1967a}. Kato's inequality, applied to a sum of correlated random variables and their expected value, provides a tighter bound compared to the widely used Azuma's inequality, leading to a higher secret key rate. Kato's inequality is presented in its general form as follows: let $\xi_1,...,\xi_n$ be a sequence of random variables which satisfies $0 \leq \xi_i \leq 1$,($i=1,2,...,n$). We define $\Lambda_k$ as the sum of these variables, i.e. $\Lambda_k = \sum_{i=1}^k\xi_i$. Let $\mathcal{F}_k$ denote the $\sigma$-algebra generated by $\{ \xi_1,...,\xi_k\}$, which is called the natural filtration of this sequence. For any $k,a \in \mathbb{R}$ and any $b$ such that $b\geq|a|$, according to Kato's inequality we have that
\begin{align}\label{upkato}
    \rm{Pr}&\left[\sum_{i=1}^kE(\xi_i|\mathcal{F}_{i-1})-\Lambda_k \geq [b+a(\frac{2\Lambda_k}{k}-1)]\sqrt{k}\right]\\
    &\leq \exp\left[\frac{-2(b^2-a^2)}{\left(1+\frac{4a}{3\sqrt{k}}\right)^2}\right],
\end{align}
in which $E(\cdot)$ denotes the expected value. If we replace $\xi_i$ with $1-\xi_i$ and $a$ with $-a$, we have another form of Kato's inequality,
\begin{align} \label{dwkato}
    {\rm Pr}&\left[\Lambda_k - \sum_{i=1}^kE(\xi_i|\mathcal{F}_{i-1}) \geq [b+a(\frac{2\Lambda_k}{k}-1)]\sqrt{k}\right]\\
    &\leq \exp\left[\frac{-2(b^2-a^2)}{\left(1-\frac{4a}{3\sqrt{k}}\right)^2}\right].
\end{align}
\par To estimate the upper bound of the observed phase error rate $\overline{E}_{p}$, we denote the detector click event in the $i$th round as a Bernoulli random variable $\xi_i$, where $i = 1,2,...,k$. If the detector clicks in the $i$th round, $\xi_i = 1$, otherwise  $\xi_i = 0$. In this way, $\Lambda_k = \sum_{i=1}^k\xi_i$ denotes the observed value of the total number of detector click events during $k$ rounds. To get the tightest upper bound of the observed phase error rate $\overline{E}_{p}$, we need to choose the optimal values for $a$ and $b$ for Eq. (\ref{upkato}) to minimize the deviation $[b+a(\frac{2\Lambda_k}{k}-1)]$. We define $\epsilon_a$ as the failure probability for estimating the upper bound, which satisfies $\epsilon_a = \exp\left[\frac{-2(b^2-a^2)}{(1+\frac{4a}{3\sqrt{k}})^2}\right]$. The solution to the optimization problem can be found in ~\cite{curras-lorenzoTightFinitekeySecurity2021a,liFinitekeyAnalysisCoherent2024}, which is
\begin{widetext}
    \begin{align}
        a_1 & = a_1(\Lambda_k,k,\epsilon_a)\\
        & =\frac{3\left(72\sqrt{k}\Lambda_k(k-\Lambda_k)\ln \epsilon_a-16k^{3/2}\ln^2\epsilon_a+9\sqrt{2}(k-2\Lambda_k)\sqrt{-k^2\ln\epsilon_a(9\Lambda_k(k-\Lambda_k)-2k\ln\epsilon_a)}\right)}
        {4(9k-8\ln\epsilon_a)(9\Lambda_k(k-\Lambda_k)-2k\ln\epsilon_a)}
    \end{align}
    \begin{align}
        b_1 & = b_1(a_1,k,\epsilon_a) = \frac{\sqrt{18a_1^2k-(16a_1^2+24a_1\sqrt{k}+9k)\ln\epsilon_a}}{3\sqrt{2k}}
    \end{align}
\end{widetext}
By substituting $a_1$ and $b_1$ into Eq. (\ref{upkato}), we obtain the expression for the upper bound of the expected value,
\begin{equation} \label{cmkatoup}
    \Lambda_k^{\ast}\leq \overline{\Lambda}_k^{\ast}=\Lambda_k+\Delta_1(a_1,b_1,k,\Lambda_k),
\end{equation}
in which we use $\Lambda_k^{\ast}$ to denote the expected value $\sum_{i=1}^kE(\xi_i|\mathcal{F}_{i-1})$ and $\Delta_1(a_1,b_1,k,\Lambda_k) = [b_1+a_1(\frac{2\Lambda_k}{k}-1)\sqrt{k}]$.
\par Similarly, solving the optimization problem corresponding to Eq. (\ref{dwkato}) yields the lower bound of the expected value. The solutions are
\begin{widetext}
    \begin{align}
        a_2 & = a_2(\Lambda_k,k,\epsilon_a)\\
        & =\frac{3\left(72\sqrt{k}\Lambda_k(k-\Lambda_k)\ln \epsilon_a-16k^{3/2}\ln^2\epsilon_a-9\sqrt{2}(k-2\Lambda_k)\sqrt{-k^2\ln\epsilon_a(9\Lambda_k(k-\Lambda_k)-2k\ln\epsilon_a)}\right)}
        {4(9k-8\ln\epsilon_a)(9\Lambda_k(k-\Lambda_k)-2k\ln\epsilon_a)}
    \end{align}
    \begin{align}
        b_2 & = b_2(a_2,k,\epsilon_a) = \frac{\sqrt{18a_2^2k-(16a_2^2-24a_2\sqrt{k}+9k)\ln\epsilon_a}}{3\sqrt{2k}}.
    \end{align}
\end{widetext}
Thus, we obtain the lower bound
\begin{equation}
    \Lambda_k^{\ast}\geq \underline{\Lambda_k^{\ast}}=\Lambda_k-\Delta_2(a_2,b_2,k,\Lambda_k),
\end{equation}
where $\Delta_2(a_2,b_2,k,\Lambda_k) = [b_2+a_2(\frac{2\Lambda_k}{k}-1)\sqrt{k}]$.
\par The above results are used to transform observed values into expected values. However, our protocol also requires the conversion of expected values back into observed values, a process also facilitated by Kato's inequality. Unlike the previous results, the observed value $\Lambda_k$ used for computing the optimal parameters $a_i, b_i$ (where $i=1,2$) is not directly observed and thus remains unknown.
Following the method in Ref ~\cite{curras-lorenzoTightFinitekeySecurity2021a,liFinitekeyAnalysisCoherent2024}, we set $a=0$ and the failure probability to $\epsilon_b$ in Eq. (\ref{upkato}) and Eq. (\ref{dwkato}), obtaining the following inequalities:
\begin{equation} \label{spkatoup}
    \sum_{i=1}^kE(\xi_i|\mathcal{F}_{i-1}) \leq \Lambda_k + \Delta
\end{equation}
\begin{equation}
    \sum_{i=1}^kE(\xi_i|\mathcal{F}_{i-1}) \geq \Lambda_k - \Delta
\end{equation}
where $\Delta=\sqrt{\frac{1}{2}k\ln \epsilon_b^{-1}}$.
\par We now briefly describe the process of calculating the upper bound of the phase error rate $\overline{E}_{p}$ using the above inequalities.  Without loss of generality, assuming Bob is the eavesdropper, Alice's raw key bits are divided into two parts based on her basis selection. When Alice selects the $X$ basis, the raw key bits are used to generate the secure key and estimate the bit error rate $E_b^X$. When Alice selects the $Y$ basis, the raw key bits are used to calculate the upper bound of the phase error rate $\overline{E}_{p}$. In the experiment, we denote the number of total detection events and the number of bit errors measured in the $Y$ basis as $n_Y$ and $m_Y$, respectively, and the number of total detection events measured in the $X$ basis as $n_X$. We then use the experimentally observed value $m_Y$ to calculate the upper bound of $m_Y^\prime$, which is the expected number of bit errors. Noting that this process transforms the observed value into the expected value to estimate the upper bound, thus, we use Eq. (\ref{cmkatoup}) to obtain the tightest bound. We utilize $E_b^{Y\prime} = m_Y^\prime/n_Y$ to calculate the upper bound of the expected bit error rate in the $Y$ basis $E_b^{Y\prime}$. Subsequently, Eq. (\ref{Yb2Xp}) is employed to compute the expected phase error rate in the $X$ basis $E_p^{\prime}$. Then, we can obtain the expected value of the number of phase errors $m_p^\prime$ in the $X$ basis by $m_p^\prime = E_p^{\prime}n_X$. Next, we use $m_p^\prime$ to calculate its upper bound $\overline{m}_{p}$. Noting that this process transforms the expected value into the observed number of phase errors in the $X$ basis, thus, we use Eq. (\ref{spkatoup}) to obtain the tightest bound. Then we can obtain the upper bound of the phase error rate in the $X$ basis $\overline{E}_{p}$ with the formula $\overline{E}_{p} = \overline{m}_{p}/n_X$.

\section{\label{appendix B}DETAILED EXPERIMENTAL DATA}
\par The precalibrated losses are shown in Table~\ref{tab:1}, including Cir, BSs, PMs, and PCs. ``PM-A (C)" stands for the PM used by Alice (Charlie), whose model number is iXblue MPX-LN-0.1. ``PM-B" stands for the PM used by Bob, whose model number is EOSPACE PM-5S5-10-PFA-PFA-UV. Detailed experimental data under different input values of pulse intensities $\mu$ and $X$-basis selection probabilities $p_{x}$ are depicted in Tables~\ref{tab:2 (a)} $\sim$ ~\ref{tab:2 (c)}, where $n$ denotes the number of clicks in the $X$ basis and $Y$ basis. ``Detected ABC" denotes the number of detection events when Alice, Bob, and Charlie, respectively add phases ``$A$", ``$B$", and ``$C$" on pulses. Note that, the insertion loss of the PC consists of the back-end attenuation of the PC to the SPD. {$\rm PC_{1}$} ({$\rm SPD_{1}$}) is on the branch connecting to the Cir while {$\rm PC_{2}$} ({$\rm SPD_{2}$}) is connected to the BS. 

\begin{table}[t]
    \renewcommand{\thetable}{II}
    \centering
    \caption{Efficiencies of the elements of the measurement station.}
    \label{tab:1} 
    \setlength{\tabcolsep}{27pt}
    \begin{tabular}[c]{cc}
        \toprule
	Optical devices & Insertion loss (dB)\\
	\midrule
	  Cir $2\rightarrow3$ & 0.49\\
	BS-1 & 0.72\\
        BS-2 & 0.68\\
	PM-A & 2.89\\
	PM-B & 3.01\\
	  PM-C & 2.23\\
        {$\rm PC_{1}$} & 0.33\\
        {$\rm PC_{2}$} & 0.45\\
	\bottomrule
    \end{tabular}
\end{table}
\begin{table*}[t]
    \renewcommand{\thetable}{III(a)}
    \centering
    \caption{Detailed experimental data under different input values of $\mu$ while $p_{x}$ is 0.9.}
    \label{tab:2 (a)}
    \setlength{\tabcolsep}{22pt}
    \begin{tabular*}{\textwidth}[c]{ccccccc}
        \toprule
	$\mu$ & \multicolumn{2}{c}{$9\times 10^ {-4}$} & \multicolumn{2}{c}{$8\times 10^ {-4}$} & \multicolumn{2}{c}{$7\times 10^ {-4}$} \\
         \toprule
	{$n$} & \multicolumn{2}{c}{1 095 111} & \multicolumn{2}{c}{951 473} & \multicolumn{2}{c}{843 121} \\
	\midrule
        Detector & {$\rm SPD_{1}$} & {$\rm SPD_{2}$} & {$\rm SPD_{1}$} & {$\rm SPD_{2}$} & {$\rm SPD_{1}$} & {$\rm SPD_{2}$}\\
	  Detected 000 & 95 331 & 553 & 84 372 & 486 & 72 840 & 475\\
        Detected 0$\pi$0 & 620 & 99 726 & 651 & 83 870 & 536 & 80 265\\ 
        Detected $\pi$00 & 910 & 105 578 & 771 & 90 217 & 737 & 80 262\\
        Detected $\pi$$\pi$0 & 92 196 & 1023 & 82 188 & 903 & 70 631 & 960\\
        Detected 00$\pi$ & 976 & 98 805 & 884 & 83 851 & 850 & 75 293\\
        Detected 0$\pi$$\pi$ & 94 686 & 1332 & 83 523 & 1379 & 73 857 & 1138\\
        Detected $\pi$0$\pi$ & 92 208 & 1070 & 82 221 & 990 & 68 962 & 880\\
        Detected $\pi$$\pi$$\pi$ & 960 & 101 433 & 997 & 86 326 & 807 & 78 385\\
        Detected 0$\frac{\pi}{2}$$\frac{\pi}{2}$ & 3 & 872 & 6 & 737 & 3 & 682\\
        Detected 0$\frac{3\pi}{2}$$\frac{\pi}{2}$ & 1597 & 11 & 1461 & 11 & 1317 & 8\\
        Detected $\pi$$\frac{\pi}{2}$$\frac{\pi}{2}$ & 1471 & 9 & 1340 & 6 & 1169 & 17\\
        Detected $\pi$$\frac{3\pi}{2}$$\frac{\pi}{2}$ & 10 & 872 & 8 & 783 & 3 & 691\\
        Detected 0$\frac{\pi}{2}$$\frac{3\pi}{2}$ & 668 & 10 & 505 & 7 & 457 & 5\\
        Detected 0$\frac{3\pi}{2}$$\frac{3\pi}{2}$ & 13 & 815 & 22 & 753 & 10 & 669\\
        Detected $\pi$$\frac{\pi}{2}$$\frac{3\pi}{2}$ & 15 & 1512 & 17 & 1266 & 18 & 1123\\
        Detected $\pi$$\frac{3\pi}{2}$$\frac{3\pi}{2}$ & 617 & 8 & 568 & 12 & 427 & 4\\
        Detected $\frac{\pi}{2}$0$\frac{\pi}{2}$ & 736 & 5 & 662 & 5 & 535 & 5\\
        Detected $\frac{\pi}{2}$$\pi$$\frac{\pi}{2}$ & 7 & 1329 & 7 & 1116 & 10 & 1015\\
        Detected $\frac{3\pi}{2}$0$\frac{\pi}{2}$ & 20 & 1638 & 17 & 1433 & 16 & 1179\\
        Detected $\frac{3\pi}{2}$$\pi$$\frac{\pi}{2}$ & 929 & 19 & 863 & 16 & 686 & 5\\
        Detected $\frac{\pi}{2}$0$\frac{3\pi}{2}$ & 19 & 1280 & 17 & 1173 & 17 & 1024\\
        Detected $\frac{\pi}{2}$$\pi$$\frac{3\pi}{2}$ & 1588 & 19 & 1333 & 24 & 1213 & 19\\
        Detected $\frac{3\pi}{2}$0$\frac{3\pi}{2}$ & 842 & 8 & 779 & 12 & 573 & 11\\
        Detected $\frac{3\pi}{2}$$\pi$$\frac{3\pi}{2}$ & 14 & 1100 & 17 & 913 & 10 & 848\\
	\bottomrule
    \end{tabular*}
\end{table*}
\begin{table*}[p]
    \renewcommand{\thetable}{III(b)}
    \centering
    \caption{Detailed experimental data under different input values of $\mu$ while $p_{x}$ is 0.8.}
    \label{tab:2 (b)}
    \setlength{\tabcolsep}{22pt}
    \begin{tabular*}{\linewidth}[c]{ccccccc}
        \toprule
	$\mu$ & \multicolumn{2}{c}{$9\times 10^ {-4}$} & \multicolumn{2}{c}{$8\times 10^ {-4}$} & \multicolumn{2}{c}{$7\times 10^ {-4}$} \\
         \toprule
	{$n$} & \multicolumn{2}{c}{1 102 508} & \multicolumn{2}{c}{968 713} & \multicolumn{2}{c}{845 563} \\
	\midrule
        Detector & {$\rm SPD_{1}$} & {$\rm SPD_{2}$} & {$\rm SPD_{1}$} & {$\rm SPD_{2}$} & {$\rm SPD_{1}$} & {$\rm SPD_{2}$}\\
	  Detected 000 & 64 617 & 378 & 57 170 & 360 & 49 161 & 288\\
        Detected 0$\pi$0 & 432 & 70 000 & 307 & 61 633 & 285 & 51 932\\ 
        Detected $\pi$00 & 687 & 74 842 & 553 & 66 464 & 432 & 56 295\\
        Detected $\pi$$\pi$0 & 66 196 & 734 & 57 695 & 776 & 50 077 & 497\\
        Detected 00$\pi$ & 882 & 73 278 & 722 & 64 648 & 707 & 57 275\\
        Detected 0$\pi$$\pi$ & 68 275 & 1105 & 59 926 & 723 & 53 385 & 868\\ 
        Detected $\pi$0$\pi$ & 69 821 & 910 & 61 279 & 877 & 54 250 & 1039\\
        Detected $\pi$$\pi$$\pi$ & 752 & 68 463 & 555 & 60 641 & 702 & 53 639\\
        Detected 0$\frac{\pi}{2}$$\frac{\pi}{2}$ & 48 & 4877 & 26 & 4448 & 22 & 3789\\
        Detected 0$\frac{3\pi}{2}$$\frac{\pi}{2}$ & 5666 & 36 & 4783 & 30 & 4239 & 40\\
        Detected $\pi$$\frac{\pi}{2}$$\frac{\pi}{2}$ & 5065 & 46 & 4594 & 37 & 3892 & 45\\
        Detected $\pi$$\frac{3\pi}{2}$$\frac{\pi}{2}$ & 31 & 3880 & 25 & 3345 & 31 & 3098\\
        Detected 0$\frac{\pi}{2}$$\frac{3\pi}{2}$ & 3171 & 63 & 2861 & 42 & 2410 & 41\\
        Detected 0$\frac{3\pi}{2}$$\frac{3\pi}{2}$ & 75 & 5032 & 59 & 4341 & 48 & 3839\\
        Detected $\pi$$\frac{\pi}{2}$$\frac{3\pi}{2}$ & 51 & 4113 & 48 & 3609 & 52 & 3293\\
        Detected $\pi$$\frac{3\pi}{2}$$\frac{3\pi}{2}$ & 3907 & 54 & 3349 & 49 & 3005 & 68\\
        Detected $\frac{\pi}{2}$0$\frac{\pi}{2}$ & 4690 & 41 & 4155 & 48 & 3583 & 33\\
        Detected $\frac{\pi}{2}$$\pi$$\frac{\pi}{2}$ & 24 & 3908 & 18 & 3459 & 20 & 2944\\
        Detected $\frac{3\pi}{2}$0$\frac{\pi}{2}$ & 40 & 4123 & 43 & 3597 & 41 & 3124\\
        Detected $\frac{3\pi}{2}$$\pi$$\frac{\pi}{2}$ & 4407 & 62 & 3816 & 63 & 3468 & 64\\
        Detected $\frac{\pi}{2}$0$\frac{3\pi}{2}$ & 58 & 5130 & 57 & 4680 & 89 & 3930\\
        Detected $\frac{\pi}{2}$$\pi$$\frac{3\pi}{2}$ & 4094 & 59 & 3599 & 55 & 3088 & 59\\
        Detected $\frac{3\pi}{2}$0$\frac{3\pi}{2}$ & 4998 & 107 & 4275 & 107 & 3787 & 125\\
        Detected $\frac{3\pi}{2}$$\pi$$\frac{3\pi}{2}$ & 75 & 4755 & 72 & 4031 & 71 & 3574\\
	\bottomrule
    \end{tabular*}
\end{table*}
\begin{table*}[p]
    \renewcommand{\thetable}{III(c)}
    \centering
    \caption{Detailed experimental data under different input values of $\mu$ while $p_{x}$ is 0.7.}
    \label{tab:2 (c)} 
    \setlength{\tabcolsep}{22pt}
    \begin{tabular*}{\linewidth}[c]{ccccccc}
        \toprule
	$\mu$ & \multicolumn{2}{c}{$9\times 10^ {-4}$} & \multicolumn{2}{c}{$8\times 10^ {-4}$} & \multicolumn{2}{c}{$7\times 10^ {-4}$} \\
         \toprule
	{$n$} & \multicolumn{2}{c}{1 102 207} & \multicolumn{2}{c}{968 120} & \multicolumn{2}{c}{856 672} \\
	\midrule
        Detector & {$\rm SPD_{1}$} & {$\rm SPD_{2}$} & {$\rm SPD_{1}$} & {$\rm SPD_{2}$} & {$\rm SPD_{1}$} & {$\rm SPD_{2}$}\\
	  Detected 000 & 41 129 & 225 & 37 040 & 205 & 31 484 & 155\\
        Detected 0$\pi$0 & 282 & 53 911 & 256 & 46 024 & 254 & 41 576\\ 
        Detected $\pi$00 & 530 & 47 769 & 429 & 41 448 & 291 & 36 086\\
        Detected $\pi$$\pi$0 & 46 439 & 558 & 41 190 & 387 & 34 844 & 354\\
        Detected 00$\pi$ & 606 & 50 815 & 523 & 44 696 & 392 & 40 961\\
        Detected 0$\pi$$\pi$ & 41 134 & 613 & 36 582 & 469 & 32 605 & 367\\ 
        Detected $\pi$0$\pi$ & 44 719 & 654 & 39 529 & 589 & 34 123 & 460\\
        Detected $\pi$$\pi$$\pi$ & 562 & 47 248 & 417 & 40 998 & 408 & 37 666\\
        Detected 0$\frac{\pi}{2}$$\frac{\pi}{2}$ & 69 & 8841 & 57 & 7781 & 63 & 6892\\
        Detected 0$\frac{3\pi}{2}$$\frac{\pi}{2}$ & 7646 & 82 & 6837 & 53 & 6245 & 71\\
        Detected $\pi$$\frac{\pi}{2}$$\frac{\pi}{2}$ & 7863 & 93 & 7048 & 58 & 5937 & 62\\
        Detected $\pi$$\frac{3\pi}{2}$$\frac{\pi}{2}$ & 61 & 7834 & 54 & 6837 & 57 & 6373\\
        Detected 0$\frac{\pi}{2}$$\frac{3\pi}{2}$ & 7859 & 170 & 7142 & 139 & 5893 & 131\\
        Detected 0$\frac{3\pi}{2}$$\frac{3\pi}{2}$ & 166 & 9127 & 133 & 7971 & 107 & 7621\\
        Detected $\pi$$\frac{\pi}{2}$$\frac{3\pi}{2}$ & 141 & 8478 & 129 & 7384 & 118 & 6584\\
        Detected $\pi$$\frac{3\pi}{2}$$\frac{3\pi}{2}$ & 7433 & 120 & 6721 & 121 & 5737 & 121\\
        Detected $\frac{\pi}{2}$0$\frac{\pi}{2}$ & 8209 & 81 & 7441 & 65 & 6393 & 64\\
        Detected $\frac{\pi}{2}$$\pi$$\frac{\pi}{2}$ & 56 & 8369 & 46 & 6983 & 44 & 6445\\
        Detected $\frac{3\pi}{2}$0$\frac{\pi}{2}$ & 133 & 9706 & 110 & 8576 & 87 & 7587\\
        Detected $\frac{3\pi}{2}$$\pi$$\frac{\pi}{2}$ & 7972 & 107 & 6910 & 80 & 6076 & 82\\
        Detected $\frac{\pi}{2}$0$\frac{3\pi}{2}$ & 146 & 8894 & 141 & 7743 & 128 & 7086\\
        Detected $\frac{\pi}{2}$$\pi$$\frac{3\pi}{2}$ & 8826 & 184 & 7718 & 118 & 6835 & 139\\
        Detected $\frac{3\pi}{2}$0$\frac{3\pi}{2}$ & 7244 & 215 & 6496 & 182 & 5487 & 149\\
        Detected $\frac{3\pi}{2}$$\pi$$\frac{3\pi}{2}$ & 139 & 7097 & 105 & 6050 & 110 & 5722\\
	\bottomrule
    \end{tabular*}
\end{table*}

\FloatBarrier


%

\end{document}